\documentclass[a4paper]{article}
\usepackage[dvips]{graphics}

\newcommand{\ybox}[2] {
\begin{center}
\resizebox{!}{#1\textheight}
{\includegraphics{#2.eps}}
\end{center}}

\begin{document}
\thispagestyle{empty}

\title{The Energy Spectrum of Cosmic Rays 
in the Range\\ 
$3 \times 10^{17} - 4 \times 10^{18}$ eV
as Measured with the Haverah Park Array}

\author{M. Ave, J. Knapp, J. Lloyd-Evans, M. Marchesini and A.A. Watson \\
\normalsize Department of Physics and Astronomy,
University of Leeds, Leeds LS2 9JT, UK}

\date{}
\maketitle

\begin{abstract}
Air showers recorded by the Haverah Park Array during the years
1974--1987 have been re-analysed. For the original estimate of the
energy spectrum, a relationship between the ground parameter $\rho$(600)
and the primary energy, as determined by Hillas in the 1970s, was used.
Here we describe the energy spectrum obtained using the QGSJET98
interaction model in the CORSIKA Monte Carlo code, together with GEANT
to simulate the detailed detector response to ground particles.  A new
energy spectrum in the range 3 $\times$ 10$^{17}$ eV to 4 $\times$
10$^{18}$ eV is presented.
\end{abstract}

\section{Introduction}

\label{intro.sec}
After many years of research the origin of cosmic rays remains
unclear. The composition of cosmic rays at high energies is not well
known and even the energy spectrum is rather uncertain. Progress is
limited primarily because cosmic rays above 10$^{15}$ eV can only be
observed, with reasonable statistics, through the extensive air showers
of secondary particles that they produce in the atmosphere.  The primary
energy and mass of the initiating cosmic ray have to be deduced from the
form and particle content of the showers. This, unfortunately, requires
use of an air shower model, based on our knowledge of electromagnetic
and hadronic particle interactions and particle decay and transport in
the atmosphere.  These models suffer from poor knowledge of the hadronic
and nuclear interactions at high energies and so the results inferred
from the measurements are, to some extent, model dependent.

At Haverah Park, UK, a 12 km$^2$ air shower array of water-Cherenkov
detectors, was operational from 1967--1987 to measure cosmic rays in
the energy range $10^{17}$ eV to $10^{20}$ eV.  Here we present a
re-analysis of data taken during 1974--1987.  At that time the energy
reconstruction of individual air shower events and the energy spectrum
were obtained with a shower model of Hillas \cite{hillas}, that gave an
empirical description of high-energy interactions.

In recent years a variety of more sophisticated models for hadron
production, based on Gribov-Regge theory, have become available which
attempt to include quantum chromodynamics in a consistent way.  The
QJSJET98 model (Quark-Gluon-String model with jets) \cite{qgsjet} for
high energy interactions, often used within the CORSIKA code
\cite{corsika} developed for Monte Carlo calculations of air showers,
has been rather successful in describing consistently a variety of
experimental results over a wide energy range.

Tools for detailed simulation of the response of the detectors to
particles of different type, energy and impact angle have also become
available, e.g. GEANT \cite{geant}, that allow us to study details of
the measuring process in a manner inconceivable 25 years ago.  The aim
of this paper is to present a re-analysis of Haverah Park data on the
energy spectrum using the QGSJET98 model in the CORSIKA code and with
GEANT simulations.  In the following paper, in this volume, we present
also new results on the cosmic ray mass composition \cite{masspaper}.

\section{The Haverah Park Array}
\label{array1}

\begin{figure}[b]
\ybox{0.35}{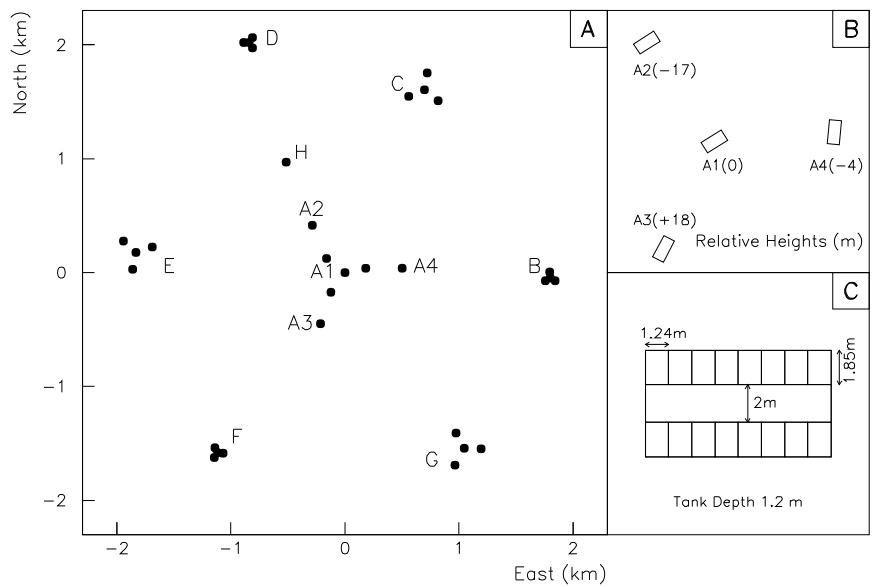}
\caption{\small Layout of the Haverah Park Array. A) The whole array. B)
The orientations and relative heights of the detector huts A1--A4. C)
The arrangement of water tanks within one of the four main A-site huts.}
\label{array}
\end{figure}

The Haverah Park extensive air shower array was situated near Leeds, UK,
at an altitude of 220 m above sea level, corresponding to a mean
atmospheric depth of 1016 g cm$^{-2}$, at 53$^\circ$ 58$^\prime$ N,
1$^\circ$38$^\prime$ W. The particle detectors of this array were made
up of water-Cherenkov modules of two types.  The majority, which were
used throughout the experiment, were galvanised iron tanks 2.29 m$^{2}$
in area and filled to a depth of 1.2 m with water brought from a nearby
borehole in magnesium limestone.  All tanks were lined with a white
diffusing material (ICI Darvic).  The water in each tank was viewed with
an EMI photomultiplier (EMI 9618YB) with a 5 inch diameter S11 Sb/CsO
photocathode held so that it just dipped into the water.  Large area
detectors were achieved by grouping together a number of modules in huts
with roofs having a thickness less than 4 g cm$^{-2}$. The signals from
the modules were added locally within each detector hut.

The number of Cherenkov photons released in a water tank is closely
proportional to the energy deposit of the shower particles in the water.
Shower electrons and photons, with typical energies of 1--50 MeV, are
effectively stopped, whilst muons (with a typical energy of 1--2 GeV)
penetrate the tank and release a number of photons that is proportional
to their track length in the water. As most of the energy of an air
shower at ground is carried by the electromagnetic particles, the use of
deep detectors is very effective for measuring the energy flow in the
shower disc.  The densities of Cherenkov photons per unit detector area
(Cherenkov densities) were recorded in terms of the average signal from
a vertical muon (1 vertical equivalent muon = 1 vem) per square
metre. This signal corresponds to approximately 14 photoelectrons for
Haverah Park tanks. Fig.~\ref{array} shows the overall layout of the
Haverah Park Array -- a more detailed description of the array is given
in \cite{oldhp}.

The signals from 15 of the 16 tanks in one A-site hut were summed to
provide the signal used for triggering and for the density estimate. One
tank in each hut was used to provide a low gain signal, used when the
signal from the other 15 tanks is saturated.  An overall trigger was
formed if, in the central detector (A1) and in at least 2 out of the
three remaining A-sites (A2--A4), a density of $>0.3$ vem m$^{-2}$ was
recorded.  The trigger rates were monitored daily over the life of the
experiment. After correction for atmospheric pressure variations, the
trigger rates were stable to better than 5\%.

In the energy range of interest here, the detectors located in a ring of
2 km from A1 do not usually show signals above threshold.  However they
are important for the analysis because they constrain the air shower
core position inside this ring.  Relevant density information is also
obtained from three 9 m$^2$ detectors at 150 m from the center of the
array. A relatively densely-packed sub array of 1 m$^2$ detectors
surrounding A1 (not shown in Fig. \ref{array}) was not used in this
re-analysis, as these detectors ran only for 4 years.

The experimental data for each event include the densities and arrival
times of shower particles for each detector.  The times were measured
relative to the start of the pulse in A1, with an experimental
measurement error of $\approx$ 20 ns.  Typically in an event the 7
central detectors have signals recorded.  For each event the arrival
direction $\theta$, $\phi$, core position x,y, and the water-Cherenkov
signal density at a core distance of 600 m, $\rho$(600), were
obtained. In total the database comprises 84455 events in the zenith
angle range between $0^\circ$ and $45^\circ$ and with energies $>2\times
10^{17}$ eV.  For the present work these data were re-analysed using
more detailed Monte Carlo calculations and improved reconstruction
algorithms, as described below.

\section{Determination of Shower Direction, Core Position and  Shower Size}

The Haverah Park data analysis procedure has three steps: firstly, the
direction of the incoming cosmic ray is determined; secondly, the core
location is estimated from the pattern of Cherenkov densities and
$\rho$(600) is determined; thirdly, the primary energy is estimated from
the relationship between $\rho$(600) and the primary energy obtained by
simulations.  The zenith and azimuthal arrival directions are determined
by fitting a plane to the particle arrival times recorded at the
detectors A1 to A4.  The error in the zenith angle determination was
estimated from the uncertainties introduced by the timing
measurements. The absolute pointing accuracy was confirmed by
comparisons with the directions of high energy muons observed in a
magnetic spectrograph located at A1 \cite{VARIOUS} and by comparison
with estimates made with an air-Cherenkov array and with an array of
scintillator detectors centered on A1 \cite{prosder}. The zenith angle
error is adequately represented by $\sigma_{\theta}=2.5^\circ~
\times~\sec\theta$, for $0^\circ < \theta < 45^\circ$ and the azimuth
angle is always smaller than that ($\approx 1^\circ$).  We use in this
work the directions determined in the original Haverah Park analysis.

Once the arrival direction has been found, the detector coordinates are
projected into the shower plane, i.e. the plane normal to the shower
axis passing through A1.  Subsequently the core location (x,y) and
$\rho$(600) in the shower plane are obtained. For this purpose we have
adopted a shape for the lateral distribution function that was
directly measured \cite{ldf}. The lateral distribution of the
water-Cherenkov density $\rho$(r), in units of vem/m$^{-2}$, in the core
distance range r = 50--800 m is given by the function:
\begin{equation}
\rho(r) = k~r^{-(\eta+r/4000~{\rm m})} = k~ f(r)
\label{ldf}
\end{equation}
where $r$ is in metres, $k$ is a normalization constant, and the slope
parameter $\eta$ is given by $\eta=3.78-1.44~\sec \theta$.  It has been
shown \cite{masspaper} that this dependence is in good agreement with
simulation results.  Due to the small number of densities usually
available, it was not possible to fit values of $\eta$ reliably for each
individual shower.  However, the value of $\rho$(600) is only weakly
dependent on $\eta$, which is one of the reasons for the choice of
$\rho$(600) as the ground parameter from which we derive the primary
energy.  Specifically the mean value of $\rho$(600) increases
(decreases) by only 15\% if the reconstruction is made using a value of
$\eta$ which is 0.3 larger (smaller) than the predicted value. Such a
shift in the value of $\eta$ corresponds to $\pm$ 3 $\sigma$ in terms of
the intrinsic shower-to-shower fluctuations of $\eta$ \cite{coy}.  We
have ignored the dependence of $\eta$ on energy, which is measured to be
0.165 $\pm$ 0.022 per decade.
 
The core position (x,y) and the shower size are fitted by comparing the
measured densities with those predicted (from Eq. \ref{ldf}), i.e. by
minimising:
\begin{equation}
\chi^2 = \sum_{i=1}^{n} \frac{1}{\sigma_i^2} 
\left( k~f(r_i) - \rho(r_i)\right)^2 \\
\label{chisqr}
\end{equation}
where $\rho_i$ are the measured densities at detector location $i$ and
$\sigma_i$ are uncertainties related to the density measurement and
particle fluctuations. By differentiating Eq. \ref{chisqr} the value of
$k$ which minimises $\chi^2$ is :
\begin{equation}
\label{normcte}
k= \frac{\sum f(r_i)   \rho_i / \sigma_i^2}
        {\sum f(r_i)^2        / \sigma_i^2} \\
\end{equation}
The $\chi^2$ minimization technique is a special case of the maximum
likelihood method of parameter estimation. If the probability
distribution function for each observed density is a Gaussian with mean
value $\rho_i(r)$ and variance $\sigma_i^2$ then $\chi^2$ is related to
the likelihood L according to :
$$
\chi^2=-2~\ln(L) + {\rm const.}
$$
In this case, however, the weights depend on x, y, $k$ and so
Eq. \ref{normcte} is no longer strictly valid. Another constraint on the
$\chi^2$ formula is that it is applicable only if the measurement
uncertainty is well described by a Gaussian distribution. When the
number of expected particles is small it is more appropriate to use
Poissonian statistics when calculating the likelihood
function. Furthermore densities below trigger threshold or above
saturation level cannot be used in the $\chi^2$ method, but it is
straightforward to incorporate them in the likelihood method. As the
measurement uncertainties are not known sufficiently well to justify a
full maximum likelihood analysis we have adopted a compromise solution,
as described below.

For a given core position and zenith angle we assume a lateral
distribution function (with the
value of $\eta$ fixed to the mean value for that given zenith angle) and
calculate the shower size constant $k$ with Eq. \ref{normcte}. We use
only detectors with densities above threshold and below the saturation
density.  The variances $\sigma_i$ used are obtained by adding in
quadrature all the possible sources of error \cite{england}:
\begin{equation}
\sigma^2_i= \rho_i^2~F_i^2+X_i^2+\frac{\rho_i}{p_i~A_i}+\frac{\rho_i}{15~A_i} \\
\end{equation}
where the first term is the absolute measurement uncertainty
($F_i=\sigma(\rho)/\rho$ is the relative uncertainty), the second term
is a constant uncertainty in $\rho$ due to digitisation, and the third
and fourth terms correspond to the Poisson uncertainty in the number of
particles and Cherenkov photons, respectively. $F_i$ and $X_i$ are
constants which depend on the type of the detector, $A_i$ is the
effective area of the detector, and $p_i$ is the ``Poisson factor'' --
the effective number of real particles or photons that constitute the
signal of a vertical equivalent muon. The Poisson factor can be $> 1$
because of the soft electromagnetic component and falls off with
distance from the shower core because of the increasing number of muons
relative to electrons and photons at greater distances. It is also
dependent on zenith angle, as the muon to electromagnetic ratio varies
with atmospheric depth. To investigate the uncertainty due to Poissonian
fluctuations in the number of particles making up the water Cherenkov
signal (as well as that due to fluctuations in the number of
photoelectrons) signals from two adjacent detector modules were recorded
for the same shower. The value of $p_i$ could be obtained with this
investigation but limited statistics preclude an accurate determination
of the dependence of $p_i$ on zenith angle and core distance.  A general
form for $p_i$ is given by:
\begin{equation}
p_i=\mu/   \check{C} + (1-\mu/  \check{C}) N_{\rm soft} \\
\end{equation}
where $\mu/$\v{C} is the fraction of the Cherenkov signal from muons,
and $N_{\rm soft}$ is the number of electromagnetic particles needed to
make up the signal of a vertical muon. The value of $p_i$ used here was
obtained from the parameterization of $\mu/$\v{C} as function of zenith
angle and core distance in \cite{armitage} with $N_{\rm soft}$=10.

Once we have obtained the value of $k$, the predicted Cherenkov
densities are calculated using $\rho_i=k~f(r_i)$ and converted to
particle numbers by multiplying with $p_i$. The likelihood function is
calculated at this stage, including detectors above threshold and below
saturation and using Poissonian or Gaussian statistics, depending on the
particle number. A gradient search of the core position (x,y) is carried
out to minimise the likelihood function.  Once the optimum core position
and $k$ are found, the assumed lateral distribution function is modified
by varying $\eta$ around its expected value. For each value of $\eta$ a
gradient search is performed to obtain new values of $k$ and core
position. The whole procedure is repeated for varying starting values of
the core position to avoid trapping in local minima.  The minimum value
of the likelihood function for all values of $\eta$, x, y, and $k$ is
assumed to be the {\em true} minimum. The value of $\rho$(600) is then
obtained from the normalization constant $k$ and the corresponding value
of $\eta$.

Two main differences between this analysis and the original Haverah Park
algorithm should be noted: 
$p_i$ varies with zenith angle and core
distance and $\eta$ is allowed to vary in the fitting procedure, whereas
both were considered constant originally. From the total event sample we
re-analysed more than 8000 events with originally reconstructed core
positions $< 500$ m away from A1.  After the new core reconstruction a
cut at 300 m was performed to ensure 100\% trigger probability at
energies above $3\times 10^{17}$ eV for all particle types, allowing a
simple and accurate calculation of the effective area for the
reconstruction of the energy spectrum.

\begin{figure}
\ybox{0.35}{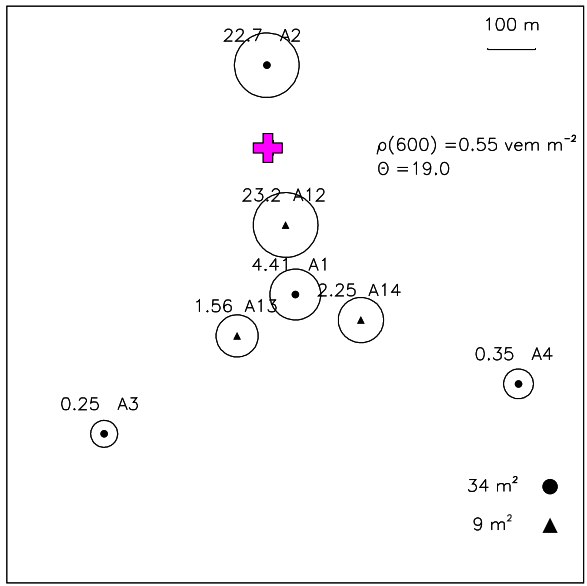}
\caption{\small Density map of a typical event recorded at the Haverah Park
Array. The area of the circles indicates the size of the Cherenkov densities,
which are shown also besides the detectors in vem m$^{-2}$. Only the inner
part of the array is displayed. The cross marks the derived core location. The
$\rho$(600) of 0.55 vem m$^{-2}$ at 19$^\circ$ of zenith corresponds to an
energy of 0.3 EeV.}
\label{event}
\end{figure}

Fig. \ref{event} shows the density map of a typical event in the shower
plane. Only the inner part of the array is displayed.

\section{Energy Calibration via Simulated Events}
\subsection{Tank Response}
\begin{figure}
\ybox{0.35}{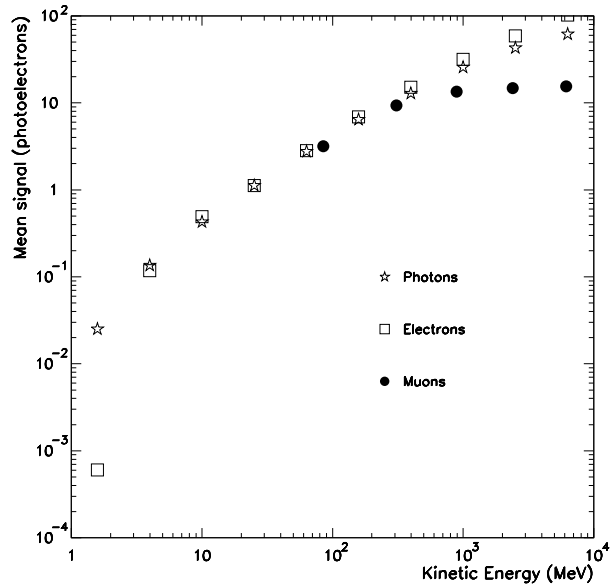}
\caption{\small The mean signal produced in a Haverah Park tank by photons,
electrons and muons, that enter a tank vertically, as a function of
energy. Low energy electrons are absorbed in the lid of the tank.  High
energy muons can penetrate the tank and deposit $\approx 240$ MeV within
it.}
\label{tank_egm}
\end{figure}

A detailed GEANT simulation of the propagation of electrons, gammas, and
muons at different zenith angles through Haverah Park tanks has been
performed.  Cherenkov photons are ray-traced until they are absorbed or
fall on the photocathode of the photomultiplier.  The wavelength dependence of light
absorption in the water and the reflectivity of the tank walls have been
taken from \cite{Wtank} but were normalised to the best estimates for
the Haverah Park tanks. The peak values for Haverah Park are 83\% for
the wall reflectivity, 15 m absorption length in the water, and 22\%
photomultiplier
quantum efficiency.  The Thorn/EMI (9618YB PMT) wavelength dependent 
quantum efficiency was taken from the manufacturer's specifications.
The results of this simulation for particles that enter the tank
vertically are summarised in Fig.~\ref{tank_egm}. The mean signal for a
typical vertical muon ($E_\mu \approx 1$ GeV) was measured to be 14
photoelectrons \cite{evans}, and this number is well reproduced by the
simulations.

By contrast, the mean energy of electrons and gamma rays in vertical
showers are around 40 and 10 MeV respectively at about 100--1000 m from
the shower core.  These calculated values are consistent with
experimental evidence from Haverah Park work \cite{towers}.  Convolving
the energy spectrum of electrons and gamma rays with the response given
in Fig.~\ref{tank_egm}, yields mean signals of 2.6 and 0.9
photoelectrons for electrons and gamma-rays, respectively, at a typical
core distance of 600 m.

Specific simulations for different zenith angles ($0^\circ$, $15^\circ$,
$26^\circ$, $40^\circ$ and $45^\circ$) and azimuth angles ($0^\circ$,
$45^\circ$, $90^\circ$) were performed to search for any change of the
signal of electromagnetic particles due to border effects. For the mean
photon signal, variations with angle were found to be less than 15\%,
and for electrons the variations are even smaller. The electromagnetic
particles usually get completely absorbed in the tanks and the output
signal is just proportional to the input particle energy.  Thus, their
contribution to the total signal at larger zenith angles is suppressed
compared to muons because of the reduction of the projected area of the
detectors. This effect can be included easily in the simulations if the
mean signal from an electromagnetic particle at a given energy is
independent of the arrival direction. The measured electromagnetic
signal can also vary within a hut because of the varying azimuthal
orientation of the tanks (see Fig.~\ref{array}). For a given zenith
angle the mean signal of the electromagnetic particles for the three
azimuth angles was simulated. Using this mean value, the error in the
electromagnetic signal when considering any azimuth is less than 7\%.

The signal due to muons, is proportional to the track length.  For a
given muon density the signal is also proportional to the tank area and
as a result the mean signal is proportional to the tank volume and
independent of the arrival direction of the shower relative to the
tank. At large zenith angles the muon signal comprises a smaller number
of muons (due to projected area) but with longer track length. Therefore
Poisson fluctuations in the total number of muons going through a tank
become more important at large zenith angles and this is accounted for
in the simulations.

To give an idea about the relative importance of the different
contributions to the total water-Cherenkov signal, we have found from
simulations that the ratio of the signal from muons to the total
water-Cherenkov signal is 0.55 (0.42) at 600 m from the shower core for
an iron (proton) shower arriving at a zenith angle of 26$^\circ$ with an
energy of 0.4 EeV.

In the following section we use the mean signals from muons, electrons
and photons at different energies and angles to produce the
water-Cherenkov lateral distribution function for individual events from Monte Carlo
simulations. The response of the tank to a particle at a given energy
and arrival direction follows a distribution that is not always
Gaussian. Although it would be more accurate to sample this distribution
in the simulations to obtain the signal produced by a particle, we use
the mean value of this distribution. We have investigated these
differences in the density range of interest for this work: the
systematic effect was found to be less than a 1\%.

\subsection{Air Shower Simulations}

The Haverah Park data were originally analysed by comparison with shower
simulations of Hillas and collaborators \cite{hillas}, using an ad-hoc
hadronic interaction model. Subsequently accelerator measurements and
the theoretical understanding of soft hadronic interactions have much
improved, and the computing power available today allows much more
detailed models to be used. At present the CORSIKA program is a standard
tool for air shower simulations, and it is used by a variety of
different experiments over the energy range from 1--10$^{12}$
GeV. CORSIKA uses EGS4 (Electron Gamma Shower code) \cite{egs} for the
simulation of the electromagnetic interactions, and features a detailed
simulation of particle propagation and decay. Different models for the
simulation of hadronic interactions are available within CORSIKA. For
this work the QGSJET98 \cite{qgsjet} model was selected.  QGSJET is
based on Gribov-Regge theory of multi-Pomeron exchange to model
high-energy soft hadronic interactions, at present the only viable
theoretical approach to model cosmic ray interactions in the atmosphere.
The QGSJET model also contains a treatment of hard collisions and the
production of mini-jets, which become dominant at very high energy
collisions, and allows a realistic simulation of nucleus-nucleus
collisions. The interaction parameters are tuned to fit a wide range of
accelerator results and to provide a secure extrapolation to higher
energies.  CORSIKA/QGSJET is fast and is able to describe a wide range
of experimental cosmic ray results: from Cherenkov telescopes at
$10^{12}$ eV \cite{qgsjet_tev}, over measurements of the hadronic,
muonic and electromagnetic shower components in the knee region
\cite{modtest}, to lateral distributions and arrival times at $\approx 5
\times 10^{17}$ eV \cite{masspaper}, and air showers at $10^{20}$ eV
\cite{qgsjet_uhecr}.

Showers initiated by primary protons and iron nuclei were simulated at
zenith angles of $0^\circ$, $15^\circ$, $26^\circ$, $40^\circ$ and
$45^\circ$ with an energy of $4 \times 10^{17}$ eV.  For a zenith angle
of $26^\circ$ showers with 0.2, 0.4, 0.8, 1.6, and 3.2 EeV were also
produced.  Statistical thinning of shower particles was applied at the
level of 10$^{-6} \times$ E$_0$ with a maximum particle weight limit of
10$^{-13} \times$ E$_0$/eV.  For each set 100 showers have been
simulated, yielding a total of 3600 showers.

The CORSIKA output was convolved with the tank response described in the
previous section.  We then fitted the resulting lateral distribution
function for muons and electromagnetic particles to the function given
in Eq. \ref{ldf} for each individual shower.  To increase the statistics
further each shower was thrown 100 times on to the array with random
core positions ranging out to 400 m from the centre of the array. The
zenith angles for these multiple Monte Carlo events is obtained by
fluctuating the zenith angle of their parent event with the
uncertainties described in the previous section.  The muon and
electromagnetic densities, including the Poissonian fluctuations, at the
location of each of the detectors is obtained with the parametrizations
described above. A check is made as to whether a simulated event
fulfills the array trigger conditions and, if so, the densities are
fluctuated according to measurement errors and recorded in the same
format as real data.  Each simulated event is then analysed with the
algorithm described in the previous section to obtain the core position
and $\rho$(600).  Events with reconstructed cores at distances larger
than 300 m from the central detector are rejected. It was found that the
number of events in which the cores are falsely reconstructed to be
outside 300 m is smaller than the number of events in which cores are
falsely reconstructed to be inside 300 m. However, this is effect is
small: at 0.4 EeV an increase of less than 1\% in the number of events
that pass the core distance cut is obtained, rising to 2\% at 3.2 EeV.

\subsubsection{Energy Calibration}

\begin{figure}
\ybox{0.3}{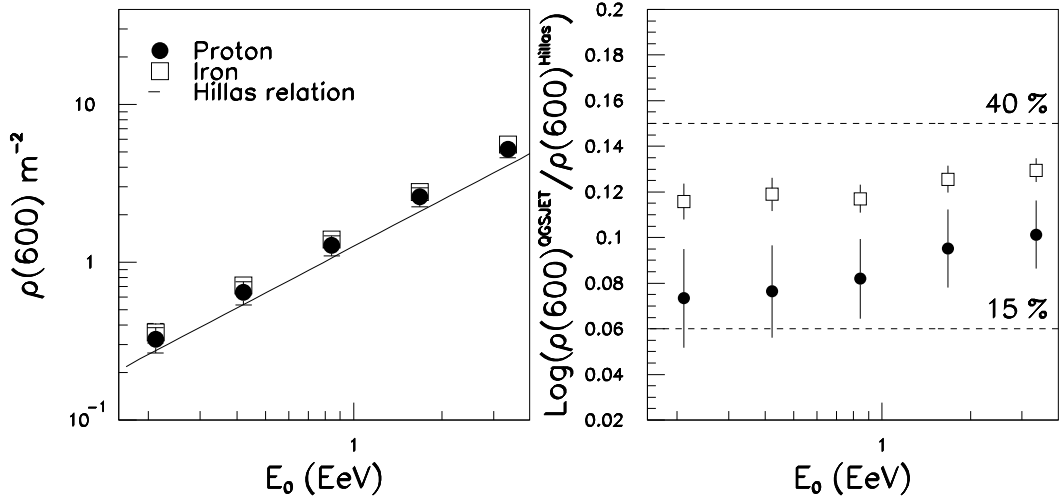}
\caption{\small Left panel: $\rho$(600) as a function of primary energy
for showers intiated by proton and iron nuclei at $\theta = 26^\circ$,
from CORSIKA/QGSJET simulations. The relation used for previous Haverah
Park analyses is also plotted as a solid line. \quad Right panel: the
ratio between $\rho$(600) obtained from CORSIKA/QGSJET and by Hillas et
al. \cite{hillas}}
\label{corr}
\end{figure}

The correlation between $\rho$(600) and the primary energy is used to
determine the energy of an event. We have studied this correlation at a
zenith angle of 26$^\circ$ since this is close to the median angle of
the showers in the data. The value of $\rho$(600) for each energy was
obtained by calculating the mean value of the reconstructed $\rho$(600)
for all simulated events generated for the corresponding set of CORSIKA
showers.  The results for proton and iron primaries are plotted in
Fig. \ref{corr}, and are compared with the result obtained by Hillas et
al. \cite{hillas} and used in previous Haverah Park analyses,
e.g. \cite{oldhp}. $\rho$(600) grows nearly linearly with $E_0$,
reconfirming that $\rho$(600) is a good energy indicator. The values of
$\rho$(600) for iron-initiated showers are $\approx $10\% higher than
those for proton-initiated showers of the same energy, almost
independent of energy.

There are significant differences between the old and the new
calibrations, which leads to $\approx$ 30\% lower reconstructed energies
at about 10$^{18}$ eV.  The right panel of Fig. \ref{corr} shows the
ratio between the old and two new calibrations, respectively. The slopes
for the proton and iron calibrations are only slightly different, see
Table \ref{r600}.

The relation between $\rho$(600) and energy can be described by
\begin{equation}
E_0 ({\rm EeV}) = C \rho(600, 26^\circ)^\gamma
\end{equation}

with $C$ and $\gamma$ given in Table \ref{r600} for proton and iron
primaries, and for the Hillas model adopted in previous works.

\begin{table}
\begin{center}
\begin{tabular}{|ccc|} 
\hline
 A  & $\gamma$  & $C$ \\
\hline
\hline
 1  & 0.998$\pm$0.008 & 0.616$\pm$0.006 \\
 56 & 1.011$\pm$0.004 & 0.574$\pm$0.003 \\
\hline 
 Hillas & 1.018        & 0.635           \\
\hline
\end{tabular}
\end{center}
\caption{\small Parameters for relation between $\rho$(600) and primary energy
for proton and Iron showers as calculated with CORSIKA/QGSJET. The
parameters calculated by Hillas are also shown. Hillas used a reference
zenith angle of 15$^\circ$, so for comparison we have converted the
parameters to a reference zenith angle of 26$^\circ$ using an
attenuation length of 760 g cm$^{-2}$.}
\label{r600}
\end{table}

\subsubsection{Attenuation Length}

\begin{figure}
\ybox{0.35}{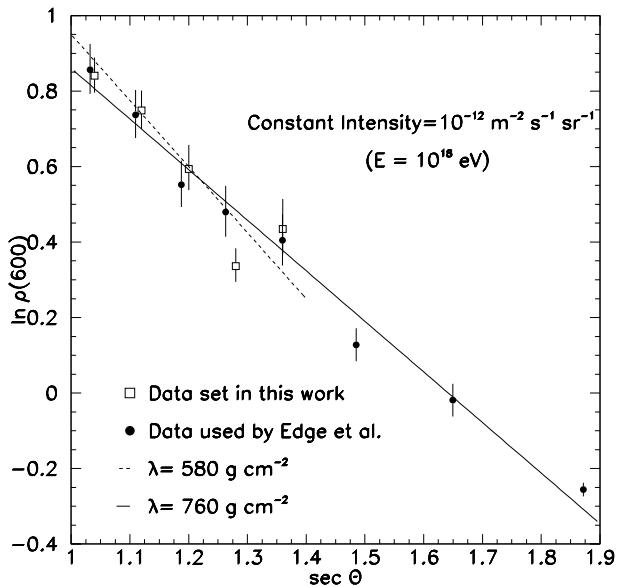}
\caption{\small Attenuation of $\rho$(600) with zenith from the constant intensity
cut method, as obtained  previously \cite{edge}, and in the present work.}
\label{atten4}
\end{figure}

Extensive air showers of different angles of incidence must be combined
to derive an energy spectrum. Here, as the calibration has been computed
for a zenith angle of 26$^\circ$, the observed density,
$\rho(r,\theta)$, at zenith angle $\theta$ and core distance $r$ is
corrected to that at $26^\circ$ using the relation:

\begin{equation}
\rho(r,26^\circ)=\rho(r,\theta)
\exp\left(\frac{X_{\rm obs}}{\lambda}(\sec\theta-\sec 26^\circ)\right) 
\label{eq-att}
\end{equation}

where $X_{\rm obs}$ is the atmospheric depth at the observation site and
$\lambda$ is the attenuation length with atmospheric depth. $\lambda$
depends on the altitude of the array and on the type of detectors used.
It can be determined from the experimental data and by simulation:
$\lambda$ is usually obtained experimentally with the ``constant
intensity cut method''. This method is based on the assumption that the
energy spectrum of cosmic rays is not dependent on the zenith angle of
observation. $\rho$(600) spectra are obtained for a set of zenith angle
bins. A horizontal slice of these spectra, at a fixed detection rate,
then delineates the attenuation of $\rho$(600) with atmospheric
depth. Fig. \ref{atten4} summarises the results from the constant
intensity cut method on Haverah Park data as obtained by Edge et
al. \cite{edge}.  Using a cut at an intensity of $10^{-12}$ m$^{-2}$
s$^{-1}$ sr$^{-1}$ (corresponding to an energy of about 10$^{18}$ eV) in
the zenith angle range 0$^\circ$ to 60$^\circ$ a value of $\lambda = 760
\pm 40$ g cm$^{-2}$ was obtained. The slope of the line was calculated
using a linear least square fit giving the points equal weight. We have
repeated this analysis for the data accumulated between 1974 and 1987,
but restricted the analysis to zenith angles in the range
0$^\circ$--45$^\circ$ and with the individual errors of the respective
points taken into account in the fit.  Our results on $\rho$(600) are
compatible with the previous Haverah Park results at smaller zenith
angles. Our estimate of the attenuation length is, however, $580 \pm 50$
g cm$^{-2}$ and is shown in Fig. \ref{atten4} as the dashed line. The
disagreement comes mainly from the different zenith angle interval used
in each calculation. The attenuation of $\rho$(600) with zenith angle is
a result of the attenuation of the electromagnetic and muonic
components; the muonic component attenuates more slowly, so when the
contribution to $\rho$(600) from the muonic component starts to dominate
the attenuation length will increase. This becomes apparent at large
zenith angles (see Fig. \ref{atten4}), and therefore this analysis is
restricted to zenith angles $<45^\circ$.

\begin{figure}
\ybox{0.35}{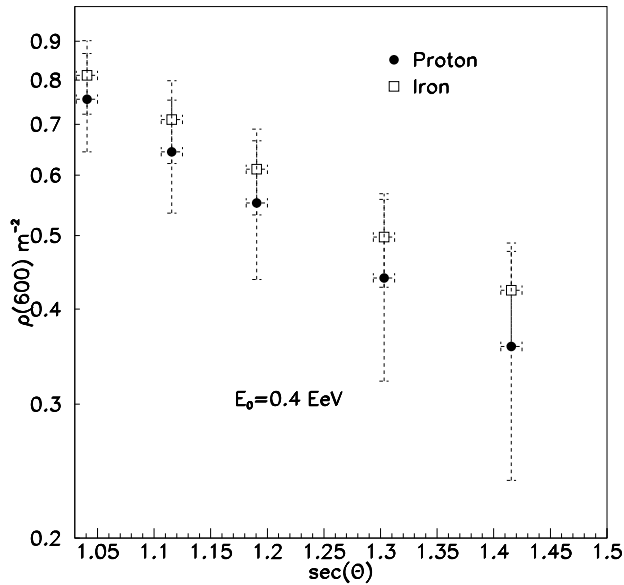}
\caption{\small Attenuation of $\rho$(600) with zenith angle for different primaries
and energies from Monte Carlo calculations. The errors bars correspond to the spread of 
 the value of $\rho$(600), to illustrate the shower fluctuations.}
\label{atten}
\end{figure}

The next step is to compare the attenuation length obtained from the
data directly with the Monte Carlo simulations. For this purpose we have
used the proton and iron-initiated showers at an energy of 0.4 EeV for
different zenith angles. The results from simulating events to obtain
mean values of the reconstructed $\rho$(600) are plotted in
Fig. \ref{atten}. The attenuation lengths obtained from this figure are
given in Table \ref{att}. They are consistent with the values from the
re-analysis of the experimental data within the statistical
uncertainties in the measurements and in the calculations.

\begin{table}
\begin{center}
\begin{tabular}{|lcc|} \hline
Energy & A & $\lambda$ (g cm$^{-2}$)\\\hline\hline
0.4 EeV & 1  & 512$\pm$50 \\
        & 56 & 581$\pm$20 \\ \hline
\end{tabular}
\end{center}
\caption{\small The attenuation length of $\rho$(600) determined for proton and iron primaries by simulation.}
\label{att}
\end{table}

\subsubsection{Energy Resolution}

The resolution of the energy reconstruction affects directly the energy
spectrum obtained from data, shifting the spectrum upward if the energy
resolution does not change with energy and shifting and changing the
slope if it does \cite{murzin}. We have estimated this energy resolution
from the simulated events for which we know the input core positions and
energies exactly.

\begin{figure}
\ybox{0.45}{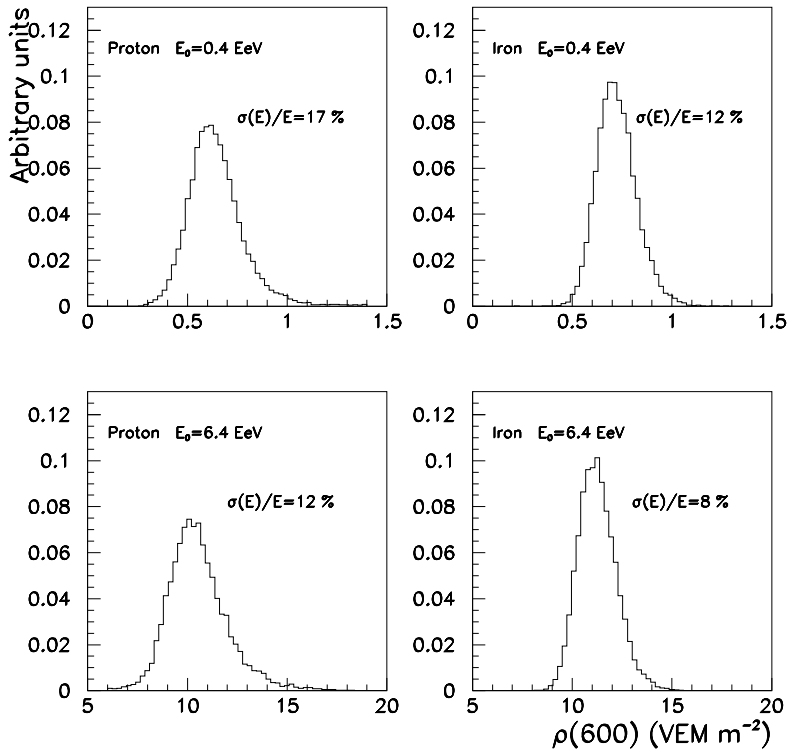}
\caption{\small Resolution of $\rho$(600) integrated over all possible zenith
angles. The attenuation lengths obtained from simultations corresponding
to each primary mass have been used.}
\label{enresol}
\end{figure}

As different zenith angles are combined to obtain the primary energy
spectrum it is necessary to take into account changes of the energy
resolution with zenith angle.  Therefore, we have used simulated events
for a fixed energy and primary mass at all zenith angles available and
have weighted them according to the number of events in each zenith
angle range. The reconstructed values of $\rho$(600) were converted to
an equivalent $\rho$(600,26$^\circ$) using the attenuation lengths given
in Table \ref{att}. Fig. \ref{enresol} shows the results of this
calculation.  The spread in the values of $\rho$(600,26$^\circ$) has
different sources: (i) experimental reconstruction errors, (ii)
intrinsic fluctuations in $\rho$(600) due to shower development, and
(iii) zenith angle errors which affect the calculation of
$\rho$(600,26$^\circ$) when using the attenuation length.

The relative widths $\sigma(\rho(600))/\rho(600)$ of the distributions
in Fig. \ref{enresol} correspond directly to the energy resolution
$\sigma(E)/E$ and are summarised in Table \ref{er}. The better
resolution for iron primaries is attributed to the smaller intrinsic
shower fluctuations. The effect of the energy resolution on the
intensities reported for the energy spectrum is less than 2\%, so no
correction has been performed.

\begin{table}
\begin{center}
\begin{tabular}{|lcc|} \hline
Energy  & A  & $\sigma$(E)/E \\\hline\hline
0.4 EeV & 1  & 17\%  \\
        & 56 & 12\%  \\ \hline
6.4 EeV & 1  & 12\%  \\
        & 56 & 8\%  \\ \hline
\end{tabular}
\end{center}
\caption{\small Energy resolution $\sigma$(E)/E for different primary energies
and masses.}
\label{er}
\end{table}

\subsubsection{Core Position}

\begin{figure}
\ybox{0.25}{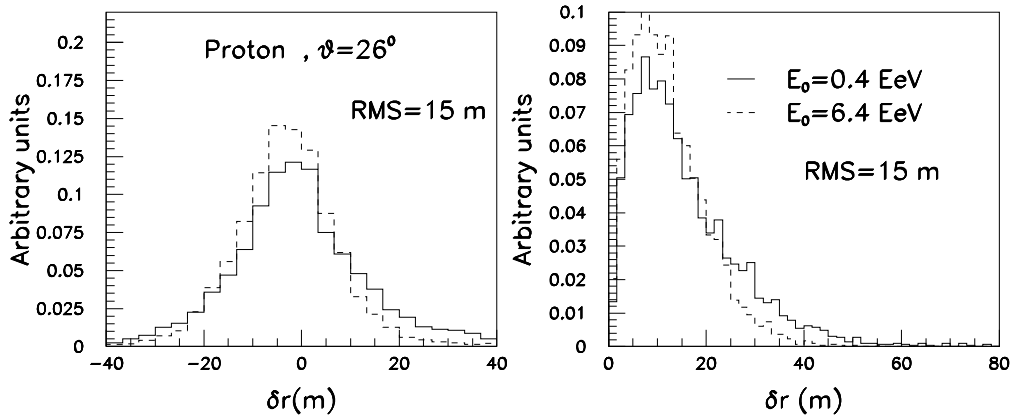}
\caption{\small Right panel: Distribution of the differences between real
(solid) and reconstructed (dashed) core position in m. Left panel:
Distribution of the differences between the distances to the central
triggering detector for real and reconstructed core position in m.}
\label{coreerr}
\end{figure}

Fig. \ref{coreerr} shows the quality of the reconstruction of the core
position for simulated proton events at two energies. The left panel
shows the distribution of differences between the distances to detector
A1 for the real and the reconstructed core positions. The right panel
shows the distribution of distances between the real and reconstructed
core position.  The widths of both distributions are $\approx$ 15 m.

\section{Energy Spectrum}

The showers recorded with the Haverah Park array, in the energy range of
interest here, provide a sufficiently large database to allow for strict
selection criteria without compromising statistical accuracy. 

All showers recorded between January 1974 and August 1987 were subject
to the following conditions.  Showers were required to have
$\rho$(600,26$^\circ$) $> 0.5$ vem/m$^{2}$. This cut removed all the low
energy events where no reliable energy reconstruction was possible. It
rejected about 90\% of the event sample.  Also shower cores had to land
within 300 m of the array centre and zenith angles had to be less than
45$^\circ$. These cuts reduced the sample by another 50\%.  The strict
selection criteria are crucial, because they ensure 100\% trigger
efficiency over the full collection area of 300 m radius, independent of
assumptions on the form of the lateral distribution function or primary
particle type.  The total on-time for these events was $3.657\times
10^8$ seconds, about 11 years. The total aperture is $7.39\times
10^{12}$ m$^2$~s~sr.

A total of 3500 events survived all selection cuts and have been used to
determine the energy spectrum.  This number shows variations up to 2\%
for the different attenuation lengths obtained in simulations, and an
increase of 30\% if the attenuation length previously used for Haverah
Park analysis ($\lambda = 760 \pm 40$ g cm$^{-2}$) was adopted.  

\begin{figure}
\ybox{0.45}{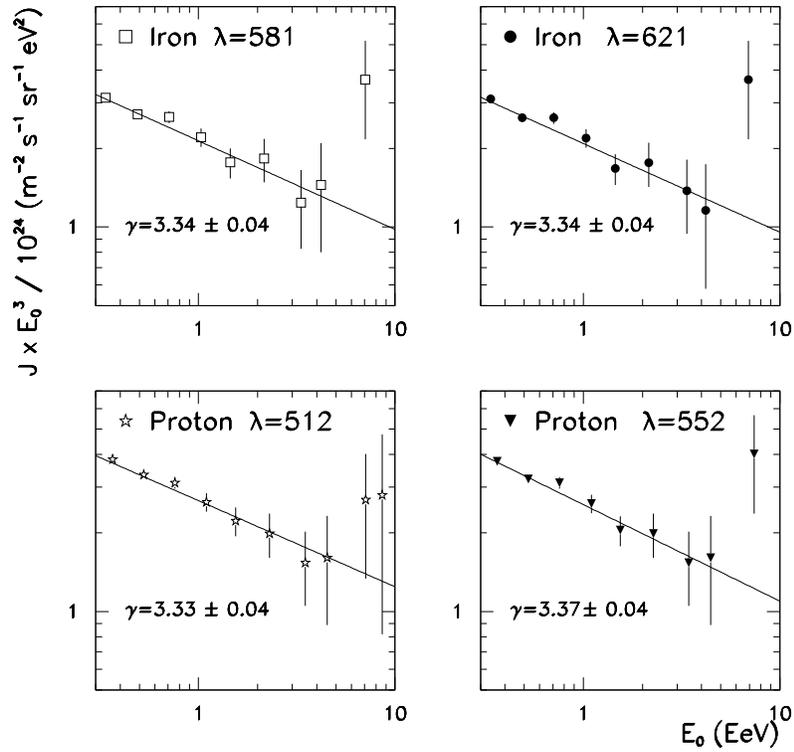}
\caption{\small Energy spectra for different assumptions on primary mass and
attenuation lengths as obtained from Haverah Park data.}
\label{graph2}
\end{figure}

The differential energy spectra obtained for different assumptions about
the primary mass composition are listed in Table \ref{proton} and
shown in Fig. \ref{graph2}.  In this graph we also show the energy
spectrum obtained if the values of $\lambda$ obtained from the
simulations are shifted by 40 g cm$^{-2}$. This shift does not affect
significantly ($< 3$\%) the final energy spectrum, showing that the
uncertainty of $\lambda$ has a negligible effect on our results.  We use
for the attenuation length the values of 512 g cm$^{-2}$ for proton and
580 g cm$^{-2}$ for iron to produce the final spectrum.

\begin{figure}
\ybox{0.35}{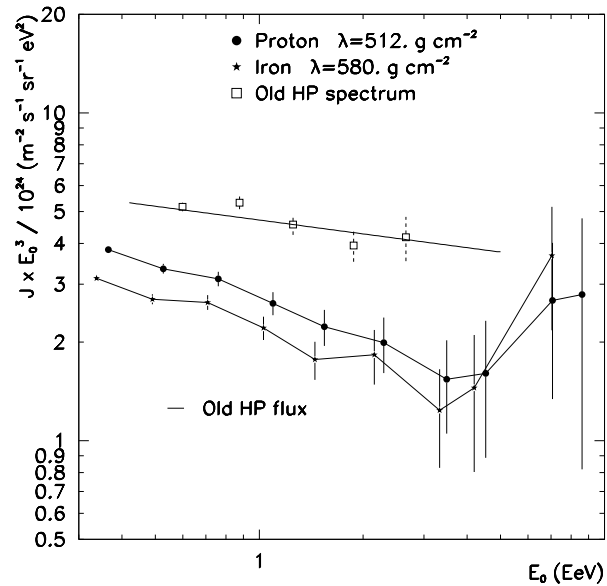}
\caption{\small Energy spectrum obtained from Haverah Park for two assumptions
of primary composition using CORSIKA/QGSJET compared with the spectra
obtained previously.}
\label{graph1}
\end{figure}

Fig. \ref{graph1} shows the differential energy spectrum obtained for
the assumption of pure proton and pure iron composition,
respectively. The Haverah Park flux reported previously \cite{oldhp} is
also shown.

\begin{table}[b]
\begin{center}
\begin{tabular}{|ccc|} 
\hline
 $\langle E\rangle$ & J$ \times 10^{30}$  & N$_{showers}$     \\
 (EeV)              & (eV$^{-1}$ m$^{-2}$ sr$^{-1}$ s$^{-1}$) & \\
\hline
\hline
0.36  & 78.5 $\pm$ 1.7  & 2052\\
0.53  & 22.8 $\pm$ 0.8  & 860 \\ 
0.76  & 7.05 $\pm$ 0.36 & 384 \\ 
1.10  & 1.97 $\pm$ 0.16 & 155 \\ 
1.54  & 0.55 $\pm$ 0.07 & 63 \\ 
2.29  & 0.16 $\pm$ 0.03 & 27 \\ 
3.49  & 0.042 $\pm$ 0.01 & 10 \\ 
4.52  & 0.014 $\pm$ 0.006 & 5 \\ 
7.07  & 0.008 $\pm$  0.004 & 4\\
8.60  & 0.0028 $\pm$  0.002 & 2 \\  
\hline
\end{tabular}
\qquad
\begin{tabular}{|ccc|} 
\hline
 $\langle E\rangle$ & J$ \times 10^{30}$  & N$_{showers}$     \\
 (EeV)              & (eV$^{-1}$ m$^{-2}$ sr$^{-1}$ s$^{-1}$) & \\
\hline
\hline
0.34  & 80.9 $\pm$ 1.8  & 1986\\
0.49  & 22.9 $\pm$ 0.8  & 815 \\ 
0.70  & 7.31 $\pm$ 0.39 & 378 \\ 
1.03  & 2.00 $\pm$ 0.17 & 150 \\ 
1.44  & 0.52 $\pm$ 0.07 & 57 \\ 
2.15  & 0.177 $\pm$ 0.03 & 28 \\ 
3.33  & 0.039 $\pm$ 0.01 & 9 \\ 
4.19  & 0.015 $\pm$ 0.007 & 5 \\ 
7.04  & 0.0124 $\pm$  0.005 & 6\\
   &  &  \\
\hline
\end{tabular}
\end{center}
\caption{\small Differential fluxes from Haverah Park, assuming a
pure proton (left) and a pure iron composition (right), respectively.}
\label{proton}
\end{table}

\begin{table}
\begin{center}
\begin{tabular}{|ccc|} 
\hline
 A & Index & J$ \times 10^{30}$  at 10$^{18}$ eV \\
   &       & (eV$^{-1}$ m$^{-2}$ sr$^{-1}$ s$^{-1}$) \\
\hline
\hline
 1 &  3.33$\pm$0.04 & 2.66$\pm$0.09 \\
 56 & 3.34$\pm$0.04 & 2.14$\pm$0.09 \\ 
 Mixture & 3.33$\pm$0.04 & 2.35$\pm$0.09 \\ 
\hline
\end{tabular}
\end{center}
\caption{\small Parameterization of the energy spectra as power law for three
assumptions of mass composition: pure proton, pure iron, and 34\% proton
66\% iron. See text for more explanations.}
\label{index}
\end{table}
\begin{figure}
\ybox{0.35}{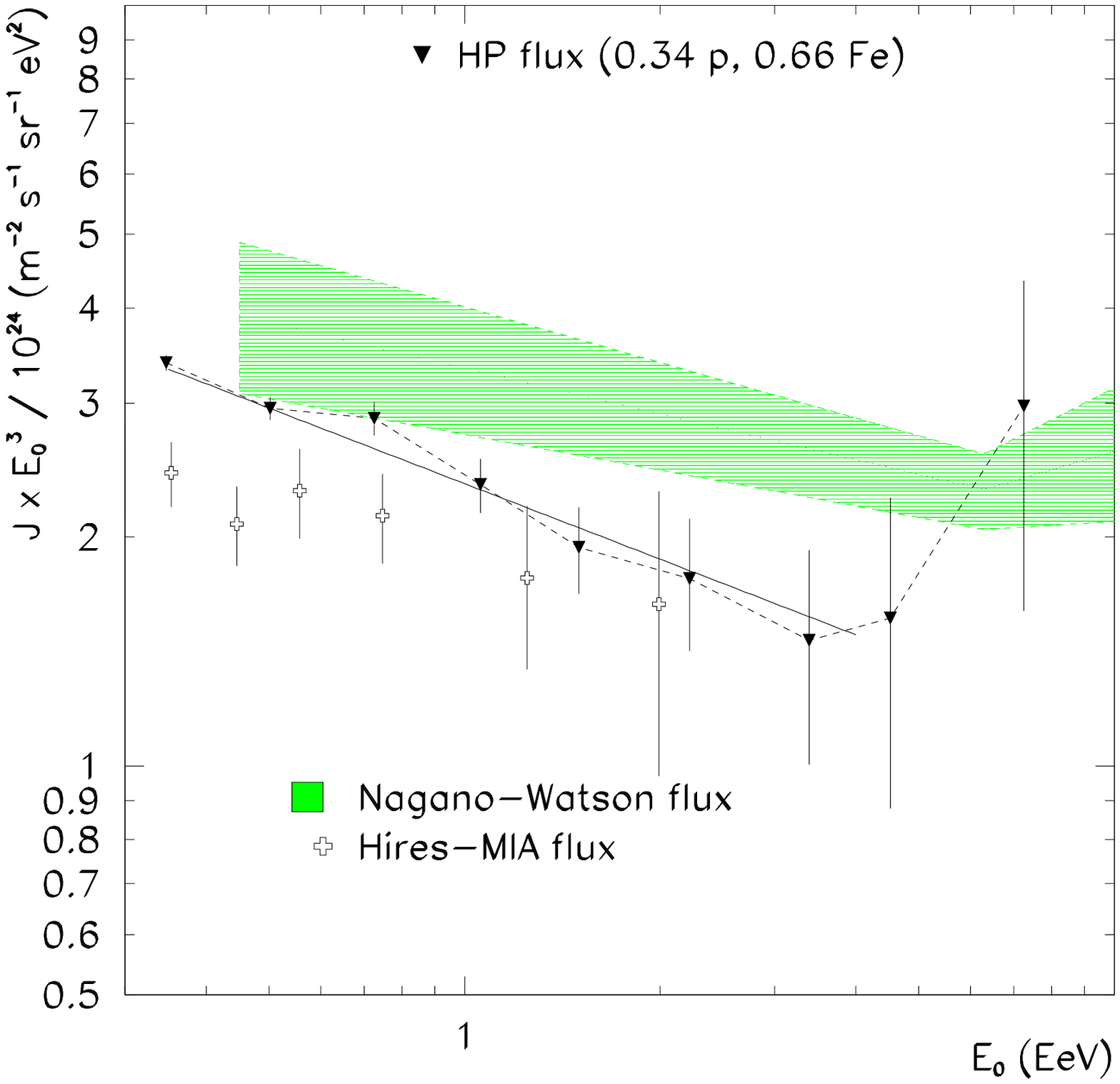}
\caption{\small Energy spectrum obtained from Haverah Park data assuming a
bi-modal composition of 34\% proton and 66\% iron as obtained in
\cite{masspaper}. Other spectra are also shown for comparison. The
continuous line shows the fit to the spectrum obtained in this work, see
table \ref{index}.}
\label{graph3}
\end{figure}

A least square fit to the experimental points in the energy range
$3\times 10^{17}$ eV to $4 \times 10^{18}$ eV is consistent with a power
law of the form dN/dE = J~(E$_0/10^{18} eV$)$^\gamma$ with the index and
normalization as given in Table \ref{index}.

It should also be noted that the difference in the $E$--$\rho$(600)
relation between different primaries induces a difference of $\approx$
20\% in the energy spectrum. In ref. \cite{masspaper} it is shown that
the lateral distribution data are in agreement with a bi-modal
composition of 34\% proton and 66\% iron in the energy range 0.3--1.0
EeV. Fig. \ref{graph3} shows the energy spectrum obtained if we adopt
this composition for the cosmic ray beam.  In this graph we also show a
recent representation of the spectrum \cite{watson} derived from Akeno
and Haverah Park data, and also the recent results obtained from the
HIRes-MIA experiment \cite{hires}.

\section{Conclusions}
The energy spectrum obtained in this analysis shows differences of up
30\% at a given energy with a recent spectrum \cite{watson} derived from
Akeno and Haverah Park data. Our results are in good agreement with the
recent results from the HIRes-MIA experiment \cite{hires} in which a
pure iron composition was assumed and with results reported at the
Hamburg conference using monocular HiRes data \cite{hiresdata}.

The spectral index after the $knee$ at 3 $\times$ 10$^{15}$ eV is
estimated to be 3.1$\pm$ 0.02 \cite{kaskade}, so a further steepening of
the primary energy spectrum between the knee and $3 \times 10^{17}$ eV
is needed to explain the spectral index derived in this work. It was
claimed in \cite{hires} that a break in the spectrum occurs at an energy
$\approx 3 \times 10^{17}$ eV. Unfortunately this is the energy
threshold of the spectrum described here.

We have re-calculated the cosmic ray flux corresponding to the 4 highest
energy events obtaining: JE$^3$=$8.27^{+6.5}_{-4.0}$ $\times$ $10^{24}$
eV$^2$~sr$^{-1}$~m$^{-2}$~s$^{-1}$ at an energy of 7.62 $\times$
$10^{19}$ eV. The average energy of these four events that were before
above 10$^{20}$ eV, is shifted a $\approx$ 30\% downwards to energies
below 10$^{20}$ eV.  The most energetic event has an energy of 8.3
$\times$ 10$^{19}$ eV.

The Haverah Park array can be considered as an early prototype of the
Auger Observatory which will employ water Cherenkov tanks of identical
depth. The steps taken to derive an energy spectrum for the Auger
Observatory are likely to be similar to those described here, and
similar problems, like the attenuation length to be used and the
$E$--$\rho$(1000) relation, will need to be addressed.

\section*{Acknowledgements}
Many people contributed to the collection of the Haverah Park data used
in this paper.  We thank them and in particular acknowledge the major
contibutions of the late Professor J G Wilson and Dr R J O Reid.  The
financial support for Haverah Park came from the UK Science and
Engineering Research Council.  M Ave and A A Watson thank the UK
Particle Physics and Astronomy Research Council for fellowships and M
Marchesini thanks the University of Leeds for a William Wright
Studentship. We are grateful to the staff of the Computing Centre of
IN2P3 in Lyon and to M. Risse for the help in producing the simulated
showers used for this analysis.

\end{document}